\documentclass[aps,prl,twocolumn,superscriptaddress]{revtex4-2}
\usepackage{graphicx}
\usepackage{amsmath,amssymb,amsfonts}
\usepackage{comment}
\usepackage{float}
\usepackage{placeins}
\usepackage{braket}
\usepackage{array}

\usepackage[colorlinks,
	linkcolor=blue,
	citecolor=blue,
	urlcolor=blue]{hyperref}

\usepackage{cleveref}

\begin{document}

\title{Cross-Correlated Force Measurement for Thermal Noise Reduction in Torsion Pendulum}

\author{Yusuke Okuma}
\email[]{okuma-yusuke319@g.ecc.u-tokyo.ac.jp}
\affiliation{Department of Physics, University of Tokyo, Bunkyo, Tokyo 113-0033, Japan}

\author{Kiwamu Izumi}
\email[]{kiwamu@astro.isas.jaxa.jp}
\affiliation{Department of Space Astronomy and Astrophysics, Institute of Space and Astronautical Science, JAXA }

\author{Kentaro Komori}
\affiliation{Research Center for the Early Universe (RESCEU), Graduate School of Science, University of Tokyo, Bunkyo, Tokyo 113-0033, Japan}
\affiliation{Department of Physics, University of Tokyo, Bunkyo, Tokyo 113-0033, Japan}

\author{Masaki Ando}
\affiliation{Research Center for the Early Universe (RESCEU), Graduate School of Science, University of Tokyo, Bunkyo, Tokyo 113-0033, Japan}
\affiliation{Department of Physics, University of Tokyo, Bunkyo, Tokyo 113-0033, Japan}

\date{\today}

\begin{abstract}
The torsion pendulum is a prevailing instrument for measuring small forces acting on a solid body or those between solid bodies. While it offers powerful advantages, the measurement precision suffers from thermal noises of the suspending wires giving rise to stochastic torque noises. 
This paper proposes a new scheme to reduce the effect of such noise by employing a double torsion pendulum and cross-correlation technique.
Cross-correlating two synthesized data streams which are composed of the rotation angles of two torsion stages, it yields the power spectral density estimate of external forces acting on the lower stage with the reduced effect from the thermal torque noises. As an example use case, we discuss the application to the study on the coupling strength of ultra light dark matter to standard model particles. Our evaluation indicates that the upper limit may be improved by an order of magnitude than the previous experiments at 2~mHz, which corresponds to about $8\times 10^{-18}~\mathrm{eV}$.

\end{abstract}

\maketitle

\section{Introduction}
Precision measurement of small forces acting on a solid body has played an indispensable role in exploring modern physics. In these measurements, the torsion pendulum has been often employed as a core instrument. Since the epoch-making measurements of the Coulomb force and gravitational constant~\cite{cavendish1798xxi} in the late 18th century, the torsion pendulum has remained as a powerful experimental tool. Nowadays, it has been successfully utilized in multiple disciplines such as Casimir force measurements~\cite{lamoreaux1997demonstration}, tests of the equivalence principle~\cite{adelberger1990testing}, verification of gravitational inverse square law \cite{hoskins1985experimental} and test of gravitational coupling between small masses~\cite{westphal2021measurement}. Moreover, using optical spring have made torsion pendulum technique more sensitive sensor \cite{wang2009improving,komori2020attonewton}.

The torsion pendulum offers advantages critical for the precision force measurement, including the provision of a low mechanical resonant frequency and decoupling of the rotational mode from the gravitational pull from the Earth~\cite{gillies1993torsion}. However, it has been known that the measurement precision for external forces is limited by thermal noise from the suspending wire(s)~\cite{saulson1990thermal,gonzalez1995brownian}. A straightforward workaround is to cryogenically cool down the apparatus to reduce the magnitude of thermal noise~\cite{newman2014measurement}. However, in practice, it could introduce a set of complex systems and therefore the design must be carried out with a great care. Furthermore, very low temperature may not be preferable in the special setups such as the ground test of inertial sensors for space application where the test environment must simulate those in orbit.

In this paper, we propose a new measurement concept for reducing the effect of thermal noise in the torsion pendulum while maintaining the ability to measure small torques or equivalently forces acting on the torsion pendulum. The core idea is to synthesize two measurement signal streams such that the cross correlated spectrum between the two provides an estimate of power spectral density (PSD) of the forces where the contamination from thermal noise is integrated away due to the lack of coherence. We determined that the scheme can be implemented in a double torsion setup as opposed to the single one. This scheme can be implemented together with a cryogenic system and other noise attenuation systems.

Albeit the drawback of losing the phase information associated with the forces being measured, it offers a way to probe the magnitude of the forces in the form of PSD. A number of experiments may benefit from this scheme, including test of gravitational inverse square law \cite{tan2016new}, gravitational coupling between small masses \cite{westphal2021measurement}, dark matter search \cite{shaw2022torsion} and the ground verification of the gravitational reference sensors for space gravitational wave antennas~\cite{amaro2017laser,luo2016tianqin, hu2017taiji,kawamura2011japanese}. 

Additionally, this paper discusses the application to the study on the coupling strength of ultra light dark matter to standard model particles, $g_{\rm{B-L}}$\cite{shaw2022torsion}, as an example. We show that our scheme can improve the sensitivity level by an order of magnitude than the previous record at 2~mHz.

\section{Setup and definitions\label{sec:generic}}
Let us assume the torques or forces being measured to be stationary stochastic. The quantity we desire to estimate in the measurement is the torques in an amplitude spectral density in Nm/$\sqrt{\textrm{Hz}}$. The torque can be subsequently converted into the effective force acting on a mass by taking the lever arm into account. 

\begin{figure}[!t]
    \centering\includegraphics[width=\columnwidth]{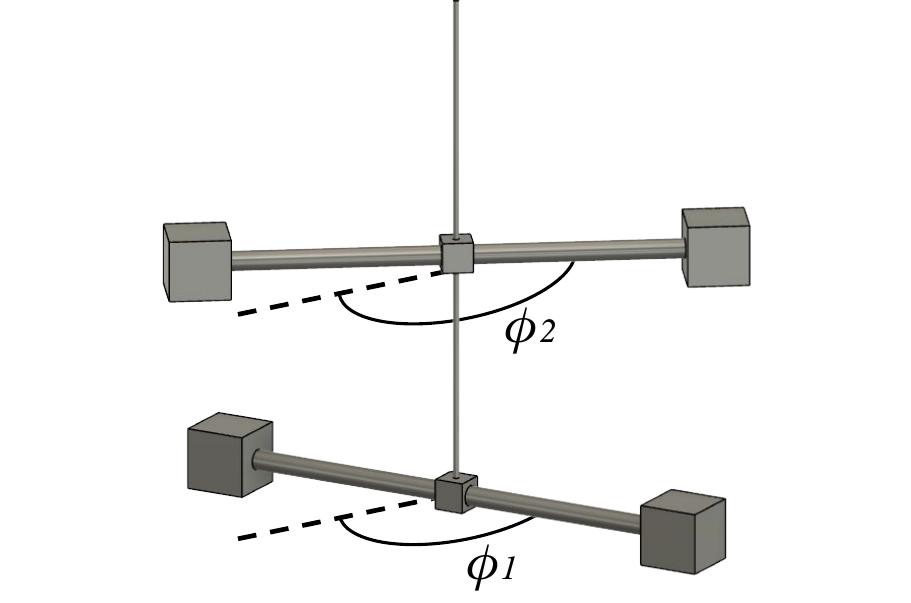}
    \caption{Illustration of two stage torsion pendulum.\label{fig_sim}}
\end{figure}

As discussed later, thermal noise from the suspending wire would not be reduced if a single-stage torsion pendulum was employed. For the reason, we begin the discussion with a two-stage torsion pendulum as shown in Fig.~\ref{fig_sim}. The lower wire suspends the first torsion mass. The upper wire and second torsion mass suspends the lower wire underneath. Each stage rotates about the wire horizontally with natural frequencies at some mHz. For simplicity, the discussion hereafter is focused on these rotational degrees of freedom only. 

The upper and lower stages receive external torques as well as thermal torque noises from the wires. The rotational angles of the two stages,  $\phi_1$ and $\phi_2$, are continuously measured and recorded.
The equations of motion for the two stages in the rotational degrees of freedom can be expressed by
    \begin{eqnarray}
    \begin{pmatrix}
        I_1\Ddot{\phi_1}\\ I_2\Ddot{\phi_2}
    \end{pmatrix}
    =
    \begin{pmatrix}
        -I_1\omega_1^2 & I_1\omega_1^2\\
        I_1\omega_1^2 & -I_2\omega_2^2-I_1\omega_1^2
    \end{pmatrix}
    \begin{pmatrix}
        \phi_1 \\ \phi_2
    \end{pmatrix}
    +
    \begin{pmatrix}
        \tau_1\\ \tau_2
    \end{pmatrix}, 
    \label{eq:eom}
    \end{eqnarray}
where $I_j$, $\omega_j$ and $\tau_j$ are the moments of inertia, resonant frequencies and torques applied to the $j$-th stage with $j\in 1, 2$. In our convention, $j=1$ corresponds to the lower stage while $j=2$ for the upper (see Fig.~\ref{fig_sim}). To reduce clutter, damping effects are implicitly incorporated in the resonant frequencies, such that $\omega_j^2$ can be later replaced in the frequency domain by $\omega_j^2\left(1+iQ_j^{-1}\right)$ with $Q_j$ a constant representing the $Q$ value for structural damping. The definitions of $\omega_j$ and $Q_j$ are chosen so that they represent the dynamics of a $j$-th torsion stage with the other stage absent or fixed.

Fourier transforming Eq.~(\ref{eq:eom}) 
, we obtain 
    \begin{eqnarray}
         \begin{pmatrix}
            \Phi_1\\ \Phi_2
        \end{pmatrix}
        =\Hat{M}
        \begin{pmatrix}
            \mathcal{T}_1\\ \mathcal{T}_2
        \end{pmatrix},
        \label{eq:eomf}
    \end{eqnarray}
where $\mathcal{T}_j\left(\omega\right)$ and $\Phi_j\left(\omega\right)$ are the Fourier domain representations of the torques $\tau_j(t)$ and rotational angles $\phi_j(t)$, respectively. $\Hat{M} \left(\omega \right)$ is a transfer matrix of the system defined by
    \begin{eqnarray}
        \Hat{M}
        &=&a
        \begin{pmatrix}
            I_2(\omega_2^2-\omega^2)+I_1\omega_1^2 & I_1\omega_1^2\\
            I_1\omega_1^2 & I_1(\omega_1^2-\omega^2)
        \end{pmatrix},\label{eq:M}\\
        \notag a&=&\frac{1}{I_1I_2\omega^4-I_1(I_1\omega_1^2+I_2\omega_2^2+I_2\omega_1^2)\omega^2+I_1I_2\omega_1^2\omega_2^2}.
    \end{eqnarray}
It is obvious that the rotational motions are consequences of  the torques which are linearly combined.

Now, even if no external forces exist, the stages rotate in stochastic manner due to the presence of torques generated by thermal noise of the two wires. Such thermal torque noise in the lower stage can be calculated by directly applying the fluctuation dissipation theorem~\cite{saulson1990thermal} as,
\begin{eqnarray}
    S^N_{\rm th}=4k_BT\Re\left\{\frac{1}{i\omega M_{11}}\right\}\quad [\rm N^2m^2/Hz]\label{eq:thFDT},
\end{eqnarray}
where $M_{11}$ is the (1,1) element of the transfer matrix $\hat M$, $k_B$ is the Boltzmann constant and $T$ the temperature.

An important remark is that Eq.~(\ref{eq:thFDT}) is consistent with the physical process that thermal torque noise of the lower wire drives both upper and lower stages in differential fashion. For this reason and later convenience, we redefine the torque vector in Eq.~(\ref{eq:eomf}) as,
\begin{equation}
    \Vec{\mathcal{T}}= N_1
    \begin{pmatrix}
    1\\0
    \end{pmatrix}
    +N_{\rm th2}
    \begin{pmatrix}
        0\\1
    \end{pmatrix}
    +N_{\rm th1}
    \begin{pmatrix}
        1\\-1
    \end{pmatrix}, \label{eq:fvector}
\end{equation}
where we have introduced three independent and uncorrelated torques, $N_1$ and $N_{{\rm th}j}$ ($j\in1,2$). $N_1$ represents an external torque acting on the lower stage with thermal noise torques removed. This is the quantity we desire to measure. On the other hand, $N_{\rm th1}$ and $N_{\rm th2}$ represent thermal torque noises from the lower and upper wires, respectively. 

One can verify that the physical process of differential torque $N_\textrm{th1}$ is incorporated in Eq.~(\ref{eq:thFDT}) by deriving the thermal torque noises from Eqs.~(\ref{eq:eomf}) and (\ref{eq:fvector}), instead. The rotation angle of lower stage is given as $\Phi_1=M_{11}N_1 + (M_{11}-M_{12})N_{\rm th1}+M_{12}N_{\rm th2}$. Then, the apparent thermal torque noise in the lower stage is expressed as
$    \frac{\Phi_1}{M_{11}} = \left(1-\frac{M_{12}}{M_{11}} \right)N_{\rm th1} + \frac{M_{12}}{M_{11} }N_{\rm th2}$
where $N_1$ is set to zero. Converting this into PSD with the PSD of each thermal torque noise being $S^N_{{\rm th}j}=4k_BTI_j\omega_j^2/\omega Q_j$, one can reach the expression identical to Eq.~(\ref{eq:thFDT}).

\section{Cross-correlated force measurement\label{sec:working}}
Plugging torque vector~(\ref{eq:fvector}) into Eq.~(\ref{eq:eomf}), one can explicitly write down the rotational angles as,
    \begin{eqnarray}
         \Phi_1 &=& M_{11}N_1+M_{12}N_{\rm th2}+(M_{11}-M_{12})N_{\rm th1},
         \label{eq:soe1}\\
         \Phi_2 &=& M_{21}N_1+M_{22}N_{\rm th2}+(M_{21}-M_{22})N_{\rm th1}.
         \label{eq:soe2}
    \end{eqnarray}
There are three unknown quantities i.e., $N_1$ and$~N_{{\rm th}j}$, while the number of equations is only two. Therefore, given the measurement of angles $\Phi_j$, one cannot solve the equation for $N_1$ without a contamination from $N_{{\rm th}j}$.

Nonetheless, we discovered a generic solution to estimate the PSD of $N_1$, namely $S^N_{11}$, independently of the other two.
Linearly combining Eqs~(\ref{eq:soe1}) and (\ref{eq:soe2}), one can obtain two synthesized data streams as,
    \begin{eqnarray}
        N_1+N_{\rm th2} &=& \frac{(M_{22}-M_{21})\Phi_1-(M_{12}-M_{11})\Phi_2}{\textrm{det}\Hat{M}}\label{eq:soef1},\\
        N_1+N_{\rm th1} &=& \frac{M_{22}\Phi_1-M_{12}\Phi_2}{\textrm{det}\Hat{M}}, \label{eq:soef2}
    \end{eqnarray}
where $\textrm{det}\hat M$ is the determinant of matrix $\hat M$. Subsequently, taking the cross correlated spectrum of two synthesized data sets, one obtains,     \begin{eqnarray}
        \notag S^N_{11}&=&\frac{1}{|\textrm{det}\Hat{M}|^2}\{M_{22}(M_{22}-M_{21})^*S^\Phi_{11}\\
        \notag&&-M_{22}(M_{12}-M_{11})^*S^\Phi_{12}-M_{12}(M_{22}-M_{21})^*S^\Phi_{21}\\
        &&+M_{12}(M_{12}-M_{11})^*S^\Phi_{22}\}, \label{eq:psd}
    \end{eqnarray}
where the asterisk denotes the complex conjugate and cross spectrum is defined as $S^\Phi_{ij}\equiv \braket{\Phi_i^*\Phi_j}$. $N_1$~and~$N_{{\rm th}j}$ are assumed to be independent of each other i.e., $S^N_{1 {\rm th}j}=S^N_{\rm th1th2}=0$ for infinite long duration of time. Therefore, one can estimate $S^N_{11}$ using the above where uncertainties for $S^N_{11}$ reduces by a factor of $k^{-1/2}$ with $k$ the number of averages. We note that since an external torque acting on the upper stage come into the system in the manner identical to $N_{\rm th2}$, it is also rejected as long as it is incoherent to the others. 

Plugging Eq.~(\ref{eq:M}) into (\ref{eq:psd}), one can explicitly write down the PSD of torque acting on the lower stage as,
    \begin{eqnarray}
        \notag S^N_{11}&=&I_1^2\omega^2(\omega^2-\omega_1^{2})S^\Phi_{11}+I_1^2\omega^2\omega_1^{2}S^\Phi_{21}\\
        \notag &&+I_1I_2(\omega^2-\omega_2^{2*})(\omega^2-\omega_1^{2})S^\Phi_{12}\\
        &&+I_1I_2(\omega^2-\omega_2^{2*})\omega_1^{2}S^\Phi_{22},
        \label{eq:sol}
    \end{eqnarray}
where $\omega_j^2$ shall now be replaced as $\omega_j^2\rightarrow \omega_j^2(1+iQ_j^{-1})$ to incorporate the damping effects.
This serves as a recipe for obtaining the cross-correlated force estimate. The rotation angles $\phi_j$ must be continuously monitored and recorded to obtain the auto- and cross-correlated spectra, $S^\Phi_{jk}$. Note that, in real experiment, one may take the real part of Eq.~(\ref{eq:sol}) to estimate the PSD as opposed to taking the absolute value. In fact, the estimation error would be smaller when the real part is evaluated.
 Besides, the mechanical properties must be characterized to obtain six parameters, namely $I_j$, $\omega_j$ and $Q_j$. An implicit assumption is that the mechanical properties are time-invariant throughout the measurements. In addition, we omit gas damping noise~\cite{Cavalleri2010} and other noises that exert torques on the lower stage. They would increase the magnitude of $S_{11}^N$ if unaddressed. Gas damping noise is found to be negligible in the numerical analyses discussed later as long as the vacuum level is at $10^{-6}$~Pa or better. 

We remark that the single-stage torsion pendulum cannot adopt this cross-correlated force estimate. In the case of a single-stage torsion pendulum, it is equivalent to set $I_2\rightarrow \infty$, resulting in det$\Hat{M}\rightarrow 0$ in Eq.~(\ref{eq:M}). This means that $N_1$ and $N_\textrm{th1}$ are degenerate unless another information is incorporated. 

One drawback in this scheme is that the phase information becomes inaccessible. The resulting estimation of external torque $n_1$ is presented in the form of PSD. Therefore, it is not possible to bring this quantity back into a set of time series data representing $n_1$.

\section{Practical limitations}\label{subsec:limitation}
The discussion so far has been restricted to an ideal situation. However, in practice, the performance of the force estimation would be deteriorated in three aspects, namely an estimation bias, residual thermal noise torques and sensing noises. While the first two effects arise due to inaccurate knowledge on the system parameters, the last adds another noise into the estimation independently of the accuracy for the system parameters. 

\begin{figure*}[!t]
    \begin{minipage}{0.45\hsize}
    \centering
    \includegraphics[width=\columnwidth]{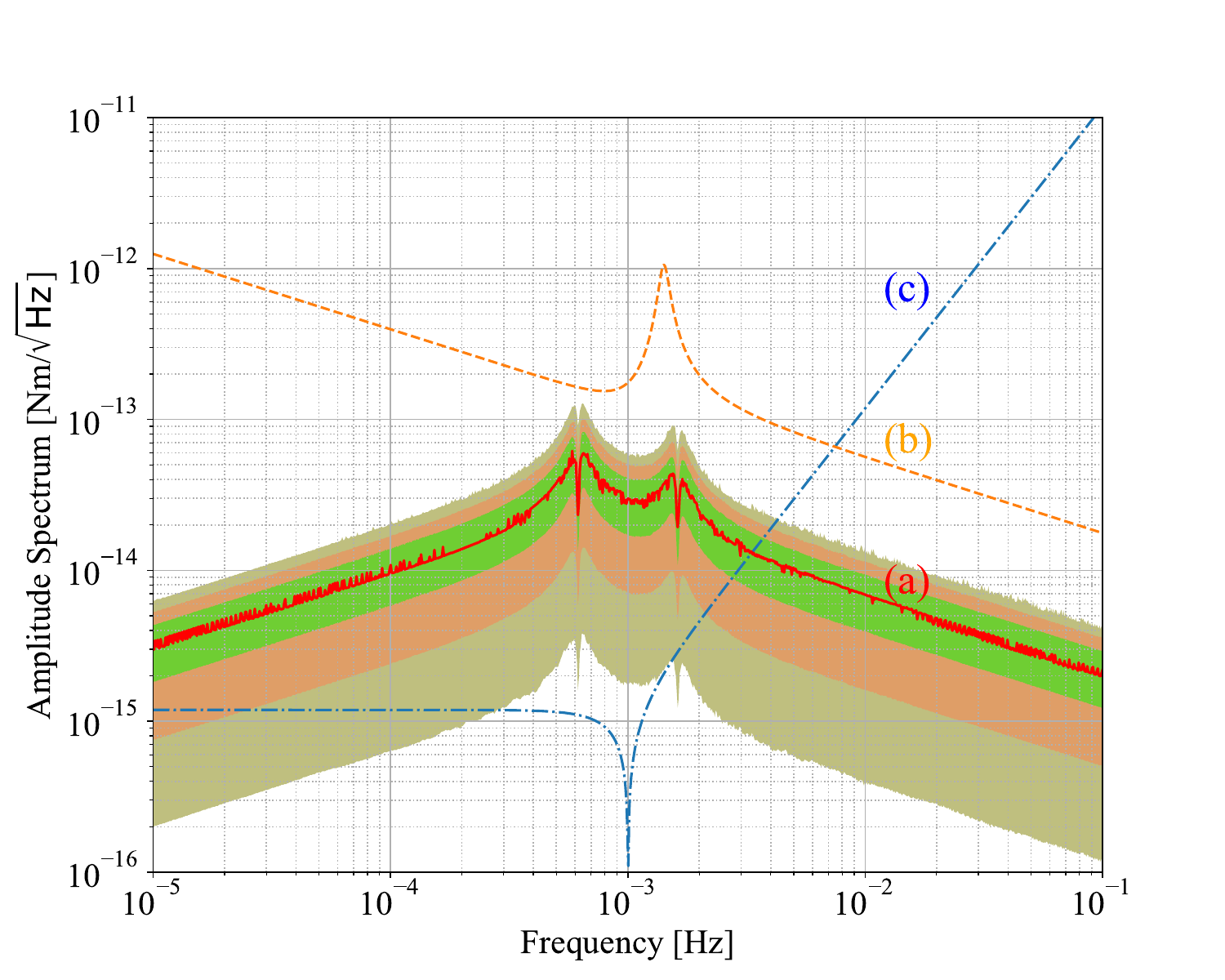}
    \end{minipage}
    \begin{minipage}{0.45\hsize}
    \centering
    \includegraphics[width=\columnwidth]{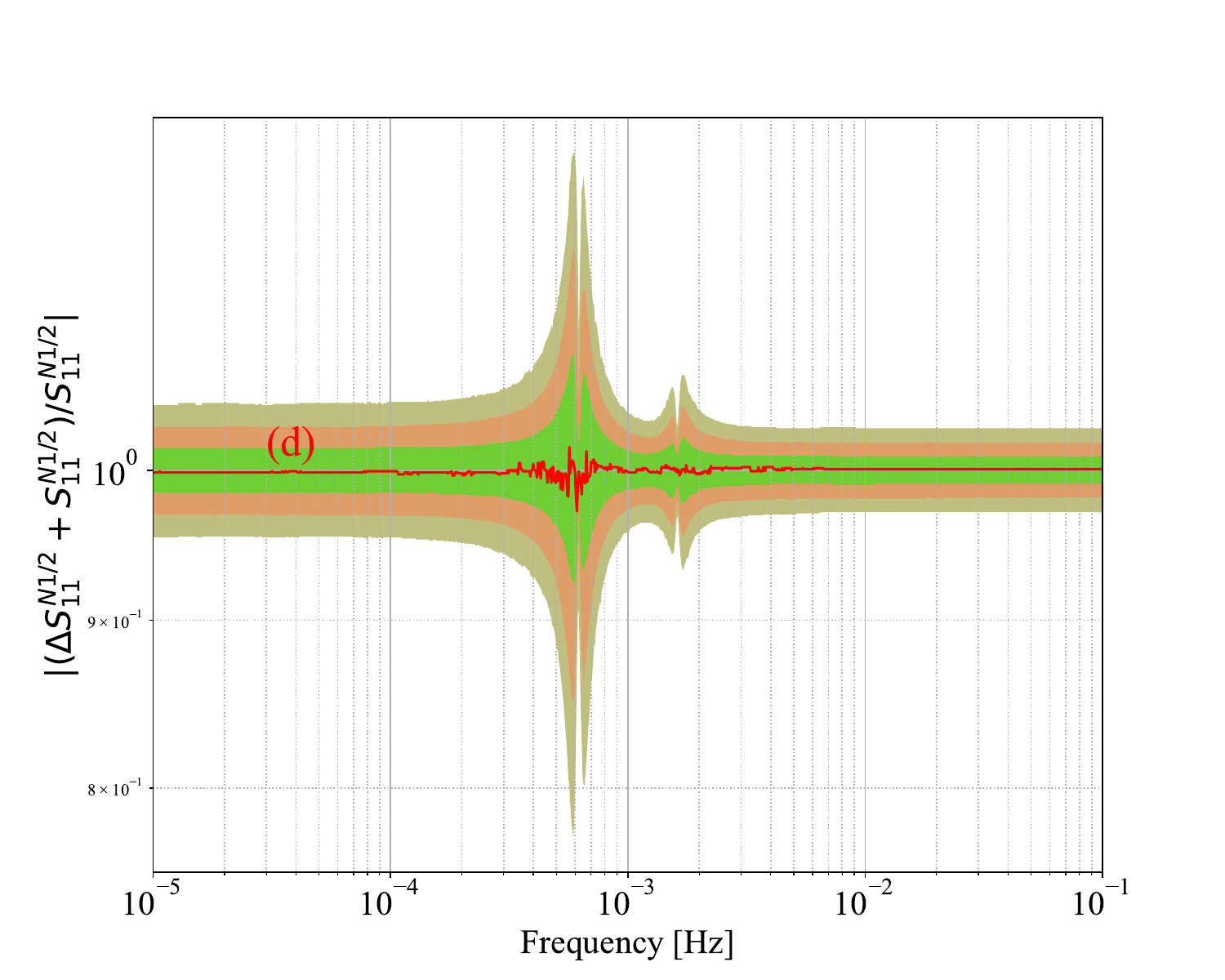}
    \end{minipage}
    \caption{(Left) Contribution of residual thermal torque and sensing noises to the torque sensitivity of lower stage.  The colored contours represent the occurrence intervals of 99.7\%, 95\% and 68\% at given frequency points. (a) Red line represents the most frequent value. (b) Dashed line shows thermal noise without cross-correlated technique using Eq~(\ref{eq:thFDT}) for comparison. (c) Dash-dotted line shows sensing noise. (Right) Relative bias in the estimate of $S^N_{11}$. Filled contours obey the same color scheme as the left panel based on the occurrence. (d) Red line represents the most frequent value. }
    \label{fig:lim}
\end{figure*}

As written in Eq.~(\ref{eq:sol}), the system parameters $I_j$, $\omega_j$ and $Q_j$ have to be known accurately. Any errors in the system parameter values would lead to an incomplete result. In fact, the resulting PSD estimate would underestimate or overestimate the true value $S^N_{11}$ and simultaneously leave residual thermal noise torques. When the system parameters are inaccurate, the right hand side in Eq.~(\ref{eq:sol}) can be expressed by,
\begin{eqnarray}
    ({\rm RHS(\ref{eq:sol})})=S^N_{11}+\Delta S^N_{11}+\Delta S^N_{\rm th1}+\Delta S^N_{\rm th2},
\end{eqnarray}
where $\Delta X$ denotes the error associated with quantity $X$. While the first term represents the true value, the second term acts as an estimation bias against the true value. The last two terms are residual thermal torque noises which would mask the true value in the spectrum.

 Table~\ref{tbl:lim} summarizes the contributions of various errors to the estimation. An error respect to the true value $\alpha$ ($\alpha\in I_j,\,\omega_j,\,Q_j$) is represented by $\delta \alpha$. The errors were expanded up to the first order and propagated to the resulting estimate in two distinct frequency regimes, namely $\omega\ll \omega_j$ and $\omega_j\ll \omega$. Similar to $Q_j$, $\delta Q_j$ are incorporated in $\delta\omega_j$ so $\omega_j\delta\omega_j$ can be replaced by $\omega_j\delta\omega_j(1+iQ_j^{-1})-\frac{i\omega_j^2\delta Q_j}{2Q_j^2}$. As seen in the table, the relative signs between the errors  matter.

To quantitatively highlight such effects, we ran a numerical simulation with a concrete set of system parameters as $T=300$~K, $I_j= 3\times 10^{-3}$~kg m$^2$, $\omega_j/(2\pi)= 10^{-3}$~Hz and $Q_j=10$. In the simulation, the system parameters were deliberately changed by $\alpha+\delta \alpha$. We adopted the Monte Carlo analysis~\cite{metropolis1949monte}, where we assumed $\delta \alpha$ to obey a gaussian distribution with zero-mean and fractional standard deviation of 0.01. In total, 100,000 realizations were computed. The system parameters were randomly sampled at each time. 

The left panel in Fig.~\ref{fig:lim} shows the effect from the residual thermal torque noises in the PSD.
As summarized in Table~\ref{tbl:lim}, the dominant contributors at frequencies higher than the resonances were identified to be errors in the moment of inertia or $\delta I_{1,2}/I_{1,2}$. This is because the mechanical responses of the system resemble a free mass response in this regime. In contrast, residual thermal torque noise was confirmed to scale as $\omega^2$ in PSD at frequencies lower than the resonances. Therefore, thermal noise is further suppressed at low frequencies.

\begin{table*}[!t]
\caption{Summary of various contributions to the resulting torque estimate.}
\label{tbl:lim}
\centering
\begin{tabular}{l|c|c}
\hline \hline
 & Low freq., $\omega \ll \omega_{1,2}$ & High freq., $\rm \omega_{1,2} \ll \omega$\\ 
\hline
Residual thermal
&$\displaystyle \frac{\omega^2}{\omega_1^{2*}}\left(-\frac{\delta I_1}{I_1}+\frac{\delta I_2}{I_2}+2\frac{\omega_2^*\delta \omega_2^*}{\omega_2^{2*}}\right)S^N_{\rm th1}+\frac{2I_1\omega^2}{I_2\omega_2^2}\frac{\omega_1\delta \omega_1}{\omega_1^2}S^N_{\rm th2}$ & $\displaystyle\left(\frac{\delta I_1}{I_1}-\frac{\delta I_2}{I_2}\right)S^N_{\rm th1}$\\
\hline
Relative uncertainty & $\displaystyle \left(\frac{\delta I_1}{I_1}+\frac{\delta I_2}{I_2}+\frac{2\omega_1\delta \omega_1}{\omega_1^2}+\frac{2\omega_2^*\delta \omega_2^{*}}{\omega^{2*}_2}\right)S^N_{11}$ & $\displaystyle\frac{2\delta I_1}{I_1}S^N_{11}$\\
\hline
Sensing noise & $-I_1I_2\omega_1^{2}\omega_2^{2*}S^\Phi_{n2}$ & $I_1^2\omega^4S^\Phi_{n1}$\\
\hline \hline
\end{tabular}
\end{table*}

The right panel in Fig.~\ref{fig:lim} shows the simulation results for the estimation bias against the true value in  $S^N_{1,1}$. The relative bias normalized by the true value is almost flat across the frequency band with amplification structures at around the resonant frequencies.

Finally, the effect of sensing noise is studied. We add moderate sensing noise of $\sqrt{ S^{\Phi}_{nj} }= 10^{-8}$~rad/$\sqrt{\textrm{Hz}}$~\cite{carbone2007upper} to the measurements of the rotational degrees of freedom. Substituting $S^\Phi_{nj}$ into $S^\Phi_{jj}$, the sensing noise after cross-correlation can be written as $I_1^2\omega^2(\omega^2-\omega_1^2)S_{n1}^\Phi+I_1I_2\omega_1^2(\omega^2-\omega_2^{2*})S^\Phi_{n2}$. $S^\Phi_{12}$ and $S^\Phi_{21}$ do not affect the sensing noise because sensing noise of $\phi_{1}$ and $\phi_2$ are independent of each other. As shown in Table~\ref{tbl:lim}, the lower moment of inertia and the lower resonant frequency, the smaller level sensing noise limit lie. This is because a small moment of inertia and resonant frequencies lead to mechanically soft system which gives a large response in rotation angles for given torques. At frequencies above $\omega_j$, sensing noise reaches the level as high as those without the cross-correlation scheme. In this region, the sensing noise scales as $\omega^4$ in PSD.

The system parameters may be fine-tuned offline to maximize the signal to noise ratio. Even if such an optimization is implemented, sensing noise would persist as a major source limiting the torque sensitivity.
Nonetheless, sensing noise can be also reduced if a pair of independent angular sensors are introduced for each stage. In fact, such a scheme has been used in the development of the gravitational reference sensor~\cite{carbone2007upper}. Therefore both thermal and sensing noises would be eliminated with the measurement for infinitely long time. In reality, the measurement time of approximately two months is required for the cross-correlated spectrum to reach the level limited by the inaccuracy of system parameters as shown in Fig(\ref{fig:lim}) above 1 mHz.

\section{Discussion}\label{sec:discussion}

\begin{figure}[!t]
    \centering
    \includegraphics[width=\columnwidth]{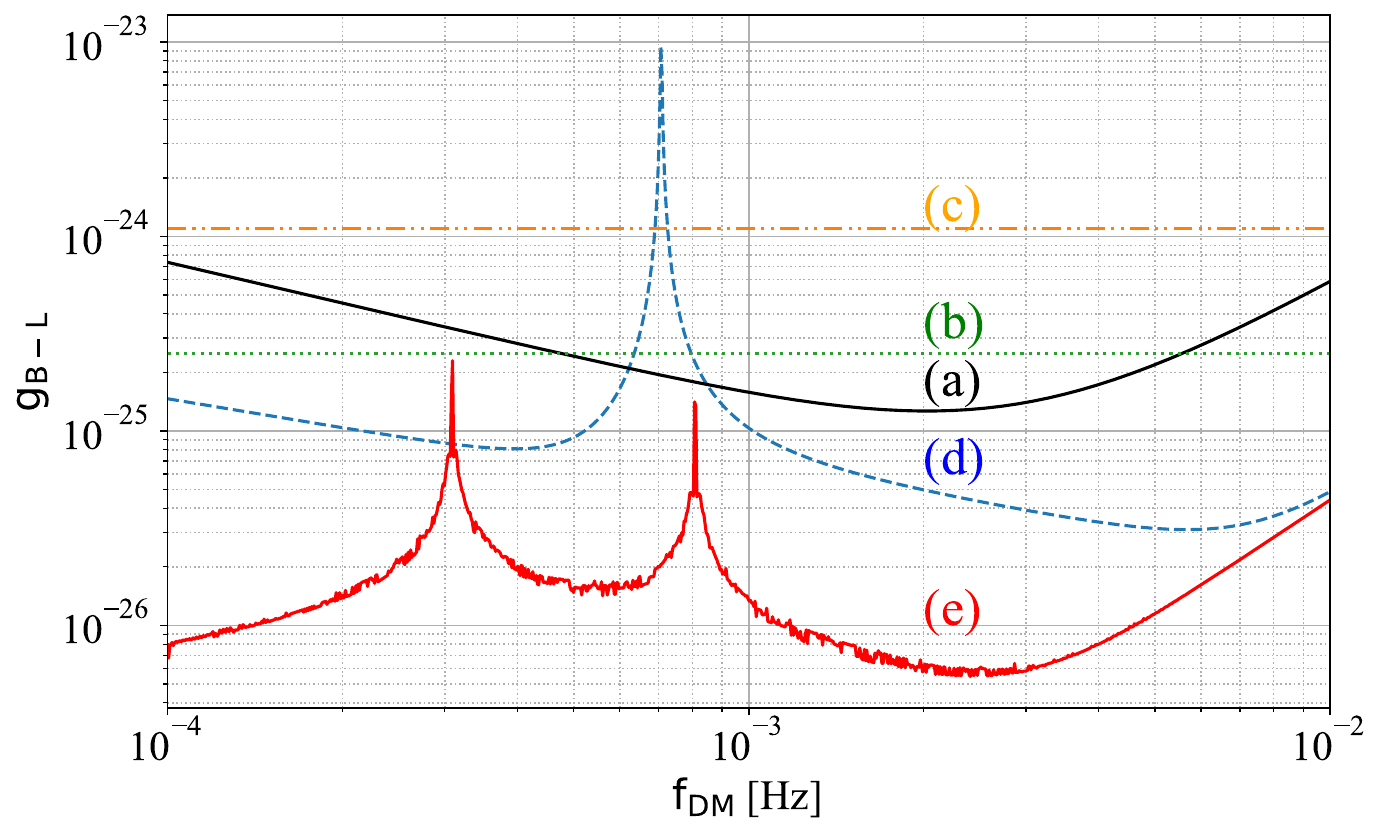}
    \caption{Expected upper limits on the coupling strength of ULDM to standard model particle. (a) Upper limits on the amplitude of field along X axis in geocentric celestial coordinates~\cite{shaw2022torsion}. (b) Upper limit by MICROSCOPE~\cite{touboul2017microscope,fayet2018microscope}. (c) Upper limit by E\"{o}t-Wash equivalence principle experiment~\cite{wagner2012torsion}. (d) Thermal noise without cross-correlation. 
    (e) Quadrature sum of the most frequent value obtained from the Monte Carlo simulation and sensing noise.}
    \label{fig:DM}
\end{figure}
One of the experiments that may benefit from the cross-correlated force measurement is the dark matter search using a torsion pendulum. Ultra light dark matter (ULDM), a candidate for dark matter, has the possibility to be composed of vector field~\cite{nelson2011dark} and could couple to B-L with B and L being baryon and lepton numbers, respectively.
Using the torsion pendulum, one can experimentally study the coupling strength of ULDM to standard model particles, $g_{\rm B-L}$. The E\"{o}t-Wash equivalence principle experiment~\cite{wagner2012torsion} limits $g_{\rm B-L}\leq 1\times 10^{-24}$ below tens of mHz. At 2 mHz, the upper limit for $g_{\rm B-L}$ is set to be less than $1\times 10^{-25}$~\cite{shaw2022torsion}, which is the most stringent limit to our best knowledge. 

In this application, a Lorentzian excess in the PSD estimate~\cite{manley2021searching}
 should be searched for. Unlike the coherent data analysis, the lack of phase information makes it difficult to determine the parameters such as phase evolution. Nonetheless, the cross-correlated force measurement is capable of improving these limits below a few mHz including the one at 2~mHz by an order of magnitude as shown in Fig.~\ref{fig:DM}.  Our expectation is derived based on~\cite{manley2021searching}. We ran another set of Monte-Carlo simulation using the system parameters $T=300$~K, $I_j= 3.78\times 10^{-5}$~kg m$^2$, $\omega_j/(2\pi)= 0.5\times 10^{-3}$~Hz, $Q_j=200$, $l=0.05$ m, $m= 19.4\times 10^{-3}$ kg, $\tau = 91.4$ days and $\Delta_{\rm B-L}$ = 0.0359, where $l$ is the arm length of torsion pendulum, $m$ is the mass of test bodies, $\tau$ is the data measurement time and $\Delta_{\rm B-L}$ the difference in the charge to-mass ratios of the two atomic species in the dipole. Uncertainties in $I_j,~\omega_j$ and $Q_j$ are assumed to obey a Gaussian distribution with zero-mean and fractional standard deviation of 0.01. We chose small $Q$ values comparing to \cite{shaw2022torsion} in which $Q =460,000$ is achieved. This is because we consider this setup also as a demonstrator for the cross-correlated force measurement. Small $Q$ values make it less stringent for reaching the thermal noise limit while it maintains the significance to dark matter search.

\section{Conclusion}\label{sec:conclusion}
This paper proposes a new scheme, the cross-correlated force measurement. The scheme utilizes the cross-correlation to overcome the sensitivity limit of thermal torque noises which would otherwise limit the sensitivity of precision force measurement in the torsion pendulum. Both qualitative and quantitative analyses for the sensitivity are presented. Inaccurate knowledge on the system parameters would limit the practical sensitivity, which invites a concrete idea for offline optimization scheme to further reduce residual thermal torque noises. The scheme can be implemented with the sensing noise reduction method~\cite{carbone2007upper}, enabling an apparatus to be significantly less sensitive to both thermal and sensing noises. As a use case, the application to the Ultra-light dark matter search is discussed. The cross-correlated force measurement is capable of improving the upper limits of $g_{\rm B-L}$ at frequencies below tens of mHz.  The scheme, however, shall be experimentally verified. 

\vspace{0.2cm}
\noindent
%\bibliographystyle{apsrev.bst}

%\bibliography{paper}
\end{document}